\documentstyle[12pt,twoside,fleqn,espcrc1,epsfig]{article}

\newcommand{\beq}{\begin{equation}}
\newcommand{\eeq}{\end{equation}}
\newcommand{\be}{\begin{equation}}
\newcommand{\ee}{\end{equation}}
\newcommand{\bea}{\begin{eqnarray}}
\newcommand{\eea}{\end{eqnarray}}
\newcommand{\ba}{\begin{array}{ccc}}
\newcommand{\ea}{\end{array}}

\def\beqn{\begin{eqnarray}}
\def\eeqn{\end{eqnarray}}

\def\Tr{ {\rm Tr} }

\newcommand{\AmS}{{\protect\the\textfont2
  A\kern-.1667em\lower.5ex\hbox{M}\kern-.125emS}}

\title{Vector condensation in QCD}

\author{K. Splittorff\address{The Niels
Bohr Institute, Blegdamsvej 17, DK-2100 Copenhagen \O, Denmark}%
\thanks{In collaboration with J.T.~Lenaghan, F~Sannino, J.J.M.~Verbaarschot, D.~Toublan, D.T.~Son, and M.A.~Stephanov}}
       
\begin{document}

\maketitle



The properties of the QCD vacuum at quark chemical potentials on the order of 
the pion mass  and at zero temperature are by now quite well understood. The
insight into this non-perturbative regime of QCD is obtained using chiral
perturbation theory at non-zero quark chemical potentials
\cite{RSSV,KSTVZ,SS,Splittorff:2000mm,KT,Lenaghan:2001sd,Splittorff:2001fy,Kogut:2001if}. 
The results apply not only to ordinary QCD but also to two color QCD and to
QCD with quarks in the adjoint representation of the color group. Some
of these realizations of QCD does not have a sign problem at finite chemical
potential and hence can be studied by standard lattice methods as well. 
This has lead to a fruitful interplay between lattice calculations and chiral perturbation
theory (see \cite{Lenaghan:2001sd} for an extensive list of references). The
central observation of all these studies is the condensation of specific goldstone   
channels. The critical chemical potentials at which the condensation occur
can be understood in terms of the Bolzman weight in the partition function
since  the effective theory at lowest order describes a non-interacting Bose
gas. For example, non-zero densities in QCD with $N_c=N_f=2$ set in for isospin
chemical potential $\mu_I>m_\pi$ or baryon chemical potential
$\mu_B>m_\pi$ \cite{SS}. This is so since in this theory there is no goldstone
mode with non-zero third component of isospin and non-zero baryon charge.

The effects at next to leading order in the chiral expansion have been studied
in two colour QCD at finite baryon chemical potential
\cite{Splittorff:2001fy}.  The renormalized theory at next to leading order
confirms that the critical baryon chemical potential is $\mu_B=m_\pi$. In addition
the critical exponents at the phase transition was calculated. Surprisingly,
they remain at the mean field values. 

The inclusion of a set of massive spin-one vectors, $A_{\nu}$, in lowest
order chiral perturbation theory for two-color QCD was studied in \cite{Duan}. 
In this case the effective Lagrangian must respect global $SU(2N_{f})$
transformations and is given by  
\begin{eqnarray}
{\cal L}_{eff}&=&f^{2}\,{\rm Tr}\left[ D_{\nu }\Sigma D^{\nu }\Sigma^{\dagger }\right]
+m_V^{2}\, {\rm Tr}\left[ A_{\nu }A^{\nu }\right]
+ h^{2}\,{\rm Tr}\left[ A_{\nu }\Sigma A^{T\nu }\Sigma^{\dagger }\right] \nonumber \\
&+& i\,sf^{2}\, {\rm Tr}\left[ A_{\nu }\Sigma D^{\nu }\Sigma^{\dagger }\right]
+f^2m_\pi^2\Tr[{\cal
  M}\Sigma+{\cal M}^\dagger \Sigma^\dagger]-\frac{1}{2g^{2}}{\rm Tr}\left[ F_{\rho \nu
}F^{\rho \nu } \right]
\ ,  \label{nr2}
\end{eqnarray}
where $\Sigma\in SU(2N_f)/Sp(2N_f)$, $A_\nu\in SU(2N_f)$, and
$ F^{\rho \nu }=\partial ^{\rho }A^{\nu }-\partial ^{\nu }A^{\rho
}-i\left[ A^{\rho },A^{\nu }\right]$. The effect of a non-zero baryon
chemical potential, $\mu_B$, in this theory was examined in
\cite{Lenaghan:2001sd}. The chemical potential enters through 
($B$ is the baryon charge matrix)
\begin{equation}
\partial_{0} \Sigma \rightarrow \partial_{0}\Sigma - i
\mu_B \left[B,\Sigma\right] \ \ \ {\rm and} \ \ \ \
\partial_{0} A_{\rho} \rightarrow \partial_{0}A_{\rho} - i
\mu_B \left[B,A_{\rho}\right] \ .
\end{equation}
From the quadratic terms in $A_\nu$ of ${\cal L}_{eff}$ one can calculate
the masses of the $(2N_{f})^2-1$ vectors as a function of $\mu_B$. An explicit
calculation for $N_f=2$ shows that \cite{Lenaghan:2001sd} three of the 15
vector masses are degenerate and independent of $\mu_B$. The latter 12
vector states mix as a function of $\mu_B$. The mixing is not complete, since
it splits into 4 sections each containing 3 vector states. Three of these
sections have the same mass spectrum. Moreover all 4 sectors have a massless
mode at $\mu_B=M_S$, where $M_S$ is the mass of the lightest vector at
$\mu_B=0$. The fact that the vector mass vanishes suggests that these 4
modes condense at this and higher values of the chemical potential. 

The method outlined here and given in detail in \cite{Lenaghan:2001sd}
directly applies to three color QCD as well as to QCD with quarks in the
adjoint representation. For instance,
one can calculate the vector masses in three colour QCD at finite isospin
chemical potential, $\mu_I$, where condensation in the $\rho$-channel 
is expected to occur at $\mu_I=m_\rho$. Note that this prediction can be examined
by means of lattice simulations.

\begin{figure}[t]
\begin{minipage}[t]{70mm}
\epsfig{file=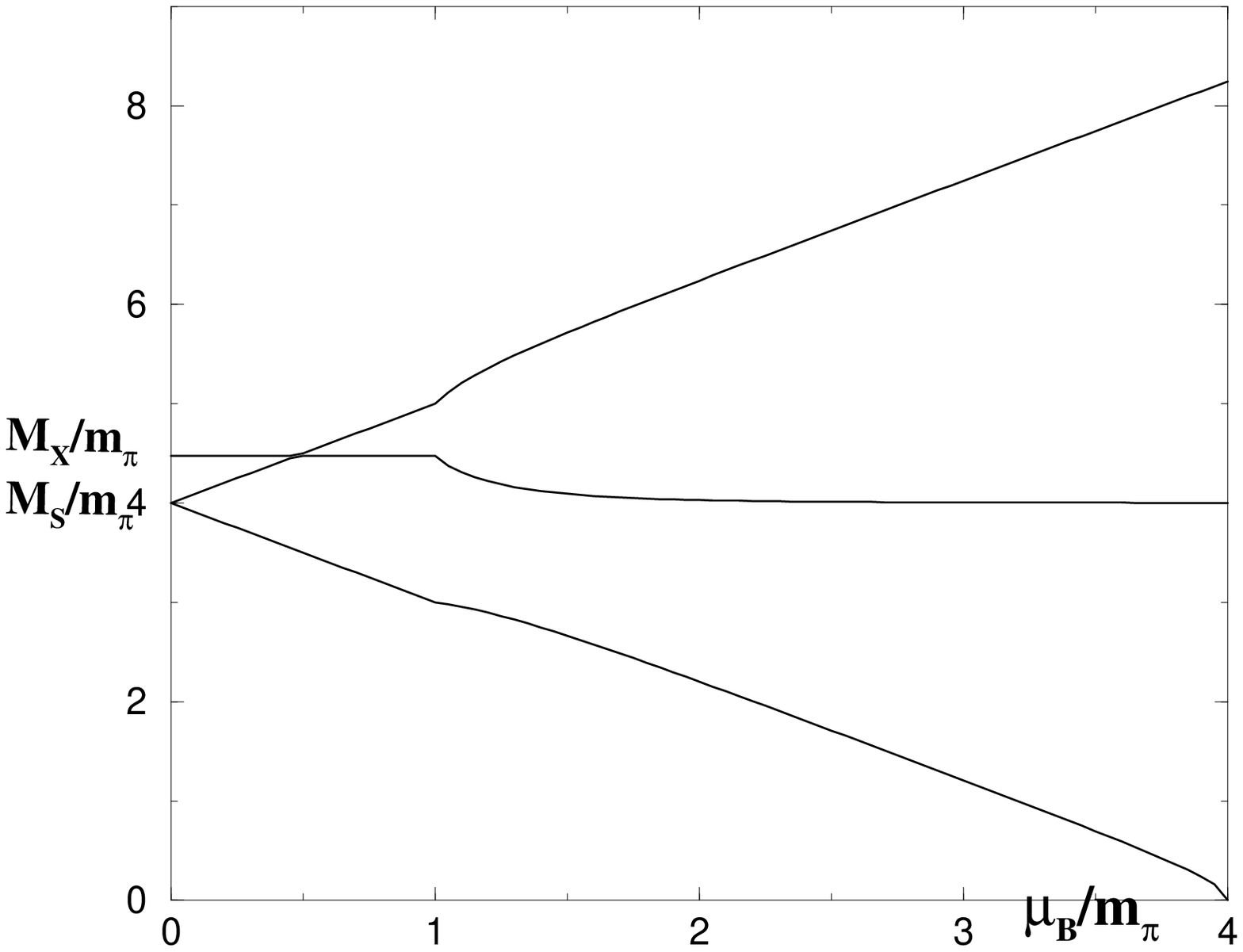,clip=,width=7.5cm}
\label{fig:3triplets}
\end{minipage}
\hfill
\begin{minipage}[t]{75mm}
\epsfig{file=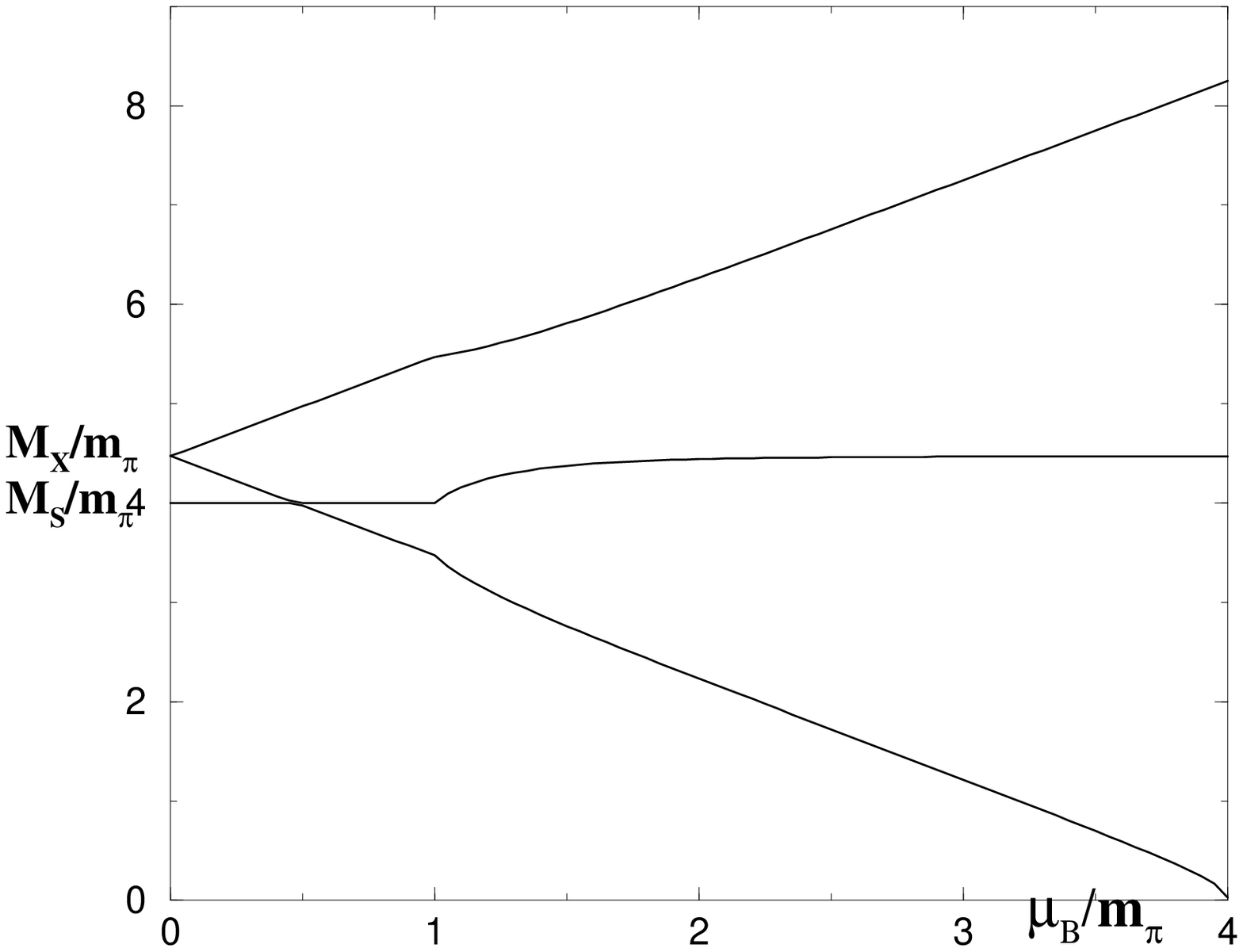,clip=,width=7.5cm}
\label{fig:A14-15-4}
\end{minipage}
\vspace{-1.1cm}

\caption{The masses of the vector states in $N_c=N_f=2$ QCD. The masses on the
  left are triply degenerate. At $\mu_B=0$ the vector mass splitting
  between vector channels associated with the broken, $X$, and unbroken, $S$,
  generators is $M_{X}^{2}-M_{S}^{2}=2 g^2 f(h+2-s)$. The parameters in
  the effective Lagrangian are taken as $(h,s,m_V,f,g)=(0,0,4m_\pi,m_\pi,1)$ .}
\end{figure}

\end{document}